# Investigating into mechanisms of high temperature strength of refractory high-entropy alloys

Sai Anandhi Seetharaman[1,*], Soumyadipta Maiti[1], Ambesh Gupta[1], Beena Rai[1]

[1]TCS Research, Tata Consultancy Services Limited, Plot No. 2 & 3, MIDC-SEZ, Rajiv Gandhi Infotech Park, Hinjewadi Phase III Pune, 411057, Maharashtra, India

* Corresponding author mail: ssai.anandhi@tcs.com

## Abstract

The yield strength plateau of two BCC refractory high entropy alloys (RHEAs) – MoNbTaVW and MoNbTaW was examined through hybrid Monte Carlo and molecular dynamics (MC/MD) simulations. By analyzing atomic diffusivities derived from vacancy formation and migration energies around the edge dislocation cores, the number of critical atomic swaps were calculated at different temperatures. Using hybrid MC/MD simulations of these critical swaps, we demonstrate that above 1400K, the stress required to move the dislocations gets saturated, indicating the effect of Dynamic Strain Ageing (DSA) via "cross core motion". Further simulations on random solid solutions (0 MC swaps) revealed a similar plateau effect at the intermediate temperatures. This was attributed to the additional athermal stress arising from lattice distortions due to solid solution strengthening. Our findings suggest that the yield strength plateau results from an interplay between the DSA-driven diffusion process and athermal stress. Specifically, the plateau emerges from DSA mechanisms in the presence of atomic diffusion, whereas in the absence of diffusion, it is governed by athermal statistical lattice distortions. This dual mechanism framework provides a comprehensive explanation for the experimentally observed Yield strength behavior in RHEAs at intermediate temperatures.

# 1. Introduction

High temperature structural materials are crucial in various industries such as aerospace, automotive, energy sector and manufacturing. These materials require high thermal stability, oxidation resistance, creep resistance to withstand extreme temperatures while maintaining their microstructural integrity and mechanical properties [1, 2]. The existing Ni-based and Co-based superalloys used for such applications provide excellent oxidation and corrosion resistance, good weldability, flexible processing, and mechanical strength due to the presence of $\gamma'$ precipitates [3]. But at high temperatures and extreme conditions, the $\gamma'$ precipitates get coarsened or dissolved, reducing their mechanical strength at high temperatures [4]. In recent years, researchers have focused on developing new materials with enhanced high-temperature strength. One such class of novel materials for high temperature applications which have drawn significant interest is High Entropy Alloys (HEAs). High entropy alloys were first perceived by Ye et al. in 2004 [5]. HEAs are also known as multi-component alloys as these alloys comprise of five or more principal elements with the composition ranging from 5 at. % to 35 at. % [6]. HEAs have garnered significant attention because their vast compositional space allows tailoring of compositions to achieve specific desired properties. Literature reports show that AlCoCrFeNiTi$_{1.0}$, AlCoCrFeNi and AlCoCrFeNiTi$_{1.5}$ HEAs exhibit specific yield strength and Young's modulus of around $0.35\ MPa.kg^{-1}.m^{-3}$ and 120 GPa compared to conventional Ni alloys which have $0.05\ MPa.kg^{-1}.m^{-3}$ and 200 GPa as specific yield strength and Young's modulus at ambient temperatures. This indicates that HEAs have much better mechanical properties than conventional Ni-based alloys [7-9].

The HEA approach was used to develop Refractory high entropy alloys (RHEAs), containing equiatomic concentrations of multi-principal refractory elements which are known for their high melting points. RHEAs are of much interest due to their remarkable stability at high temperatures, resistance to softening etc. Compared to the conventional HEAs, RHEAs have exhibited retention of yield strength at intermediate to high temperatures [7][9,10].

RHEAs were first introduced by Senkov et al. targeting aerospace and structural applications. Senkov et al. reported that the MoNbTaW RHEA forms a single solid solution exhibiting a yield strength of 1058 MPa at 25 °C, which remains as high as 552 MPa even at 800 °C [9]. Similarly, the yield strength of HfNbTaTiZr RHEA decreased with an increase in temperature from 535 MPa at 800 °C to 92 MPa at 1200 °C. Incorporating Mo or substituting Nb with Mo in equimolar HfNbTaTiZr enhanced the yield strength at 1000 ºC and 1200 ºC. Specifically, HfNbMoTaTiZr achieved yield strengths of 814 MPa and 556 MPa, while HfMoTaTiZr reached 855 MPa and 404 MPa at the same temperatures, respectively [11,12].

In case of light-weight RHEAs, the multiphase CrNbTiVZr alloy has yield strength of 1230 MPa and 615 MPa at 600 ºC and 800 ºC respectively [11][13]. Apart from these alloys, other RHEAs such as MoNbTaTiVW, MoNbTaTiW also exhibit remarkable yield strength around 753 MPa, 620 MPa even at 1000 ºC in annealed conditions [11-14].

In RHEAs, the yield strength does not decrease with an increase in temperature at intermediate temperatures (1100 K -1400 K) as observed by [15-17] which is presented in Fig.1. This yield strength plateau makes RHEAs a potential candidate for various high temperature applications.

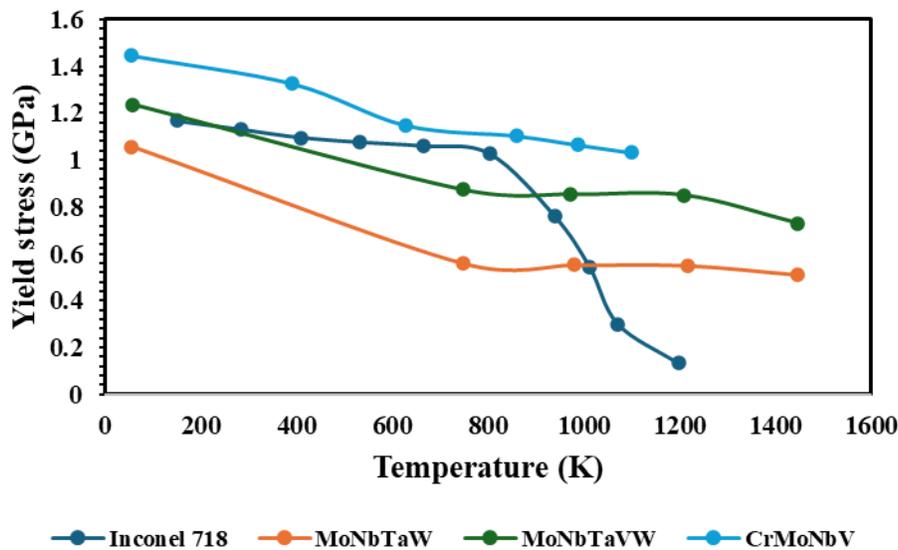

Fig.1. Yield strength (GPa) vs Temperature (K) for RHEAs, and superalloys [16].

The yield strength of BCC alloys is known to be dominated by screw dislocations as observed in TEM studies which show long straight screw dislocations punctuated by jogs/dipoles [17]. Maresca et al. predicted the CRSS of $Nb_{1-x}Mo_x$ binary alloy using the screw dislocation model developed by Statham et al. and proved that the CRSS is in particularly good agreement with Statham et al. data and experimental ones [17-18]. Both theory and experiments show that the strength of conventional BCC alloys is clearly dominated by screw dislocations.

In HEAs, dislocation dynamics are instead governed by the Nanoscale Segment Detrapping (NSD) mechanism, where both edge and screw dislocations must overcome local energy barriers under applied stress. This behavior marks a departure from the mechanisms typically observed in pure BCC elements [22]. TEM observations in $Nb_{1-x}Mo_x$ and the classic $Fe_{1-x}Si_x$, $x \leq 9\%$ alloy revealed that edge dislocations have a decreasing mobility with increasing solute content. Statham et al. stated that increase in solute contents lead to an appreciable reduction in the edge to screw mobility ratio. Similarly, Curtin et al. predicted the yield strength of $Nb_{1-x}Mo_x$ up to x = 50%. It was revealed that edge dislocations show dominance with increase in solute content and temperature as compared to screw dislocations which was consistent with earlier findings [18]. TEM studies [20] reveal screw dislocation dominance at higher plastic strains; however recent observations [21] at higher temperatures (~ 400 °C) show curved dislocations with viscous motion, suggesting a reduction in strong screw dominance. These studies show that edge dislocations dominate deformation in BCC RHEAs at intermediate and high temperatures, while screw dislocations prevail at low and room temperatures[22]. Thus, it becomes important to study deformation via edge dislocations at high temperatures to understand the underlying reason for yield strength plateau in BCC RHEAs.

The experimental procedures involved in studying the mechanical behavior of RHEAs are usually cumbersome, expensive and require assistance from skilled professionals. Also, experimental procedures come with a lot of trial - error which would lead to huge material costs and waste. To avoid this, researchers have rationally approached computational designing of RHEAs. This approach minimizes the need for experimental trials by using computational methods to establish structure-property relationships based on processing routes, with the goal of making materials development more efficient and resource-effective. There has been significant progress in the computational methods for predicting the microstructure and mechanical properties of various materials [23].

Maresca et al. developed an analytical model based on edge dislocations to predict the yield stress with temperature for MoNbTaVW and MoNbTaW RHEAs [17]. In that model, only atomic size mismatch was considered as the primary mechanism, which limited its ability to predict the yield strength plateau. As a result, the plateau was attributed to dynamic strain ageing (DSA). DSA involves rapid ageing processes that occur concurrently with material deformation, leading to enhanced strength. This phenomenon is observed as discontinuous plastic flow, evident from serrated stress-strain curves, commonly referred to as the Portevin-Le Chatelier effect. This phenomenon is usually observed in low alloy steels within the temperature range of 150-350 ºC. In this process solute atoms diffuse near to dislocations, forming atmospheres which temporarily pin them, and thereby enhancing the yield strength [24]. Recent studies show that even RHEAs show a similar type of serrated stress-strain curve. For example, Hsu et al. [25] observed that the HfNbTaTiZr system shows serrated stress strain behavior of several types over the temperature range of 400 ºC to 800 ºC. The serration behavior in materials is attributed to certain factors like temperature, pinning of solutes and strain rate. However, in RHEAs, the major contributing factor to serration is the diffusion and segregation of certain elements around the dislocation core which resists the dislocation movement leading to its high strength. This diffusion of atoms around dislocation core is known as "cross core motion" [17, 26-27].

Apart from DSA, another important factor for the yield strength plateauing in materials is attributed to athermal stress. These athermal stresses arise due to the lattice distortion from mismatches in atomic radius and elastic moduli in the random solid solution alloy [15]. According to Cornard et al. [28], the stress experienced by an edge dislocation inside a material can be split into two components, namely athermal and thermal components, respectively. When the dislocation moves through the material, the internal stress field experienced by the dislocation is in the form of two types of stress fields, i.e., long-range internal stress fields and short- range stress fields. This can be visualized as a sinusoidal-like curve with short-range stress obstacles superimposed on long-range stress fields. The energy associated with dislocations motions over long-range stress fields cannot be overcome by thermal fluctuations, whereas the short-range stress field can be overcome by combined action of applied stress and thermal fluctuations. Based on this idea, Toda-Caraballo et al. [29] developed a mathematical model to predict the yield stress without considering the thermal effects.

According to Toda-Caraballo, the yield strength plateau arises from the solid solution strengthening (SSS) due to lattice distortions in the material. The theory of solid solution strengthening started from Fleisher [30], where he studied the effect of solute atoms in strengthening of the alloys. According to the theory, the SSS is dependent on the elastic misfit and atomic size misfit. To take this approach to the next step, Labusch et al. [31] formulated the expression by considering higher concentration of solute atoms where dislocation will be constantly impeded by solute atoms.

Both approaches are applicable to binary solid solutions and do not account for temperature effects. While the Labusch et al. [31] approach aligns well with binary alloys as predicted yield strengths closely match with experimental data [32-34]. However, the yield strength values predicted for HEAs did not match with the experimental ones. This may be attributed to the complexity, multi-principal nature of HEAs and atomic mismatch caused by the solute atoms which Labusch model did not consider. In HEAs, the mobility of dislocations is consistently hindered by solutes because of the existence of over 5 principal elements, leading to continuous variations in elastic interactions throughout the lattice [35]. Based on this, Coury et al. [15] developed an analytical model where the athermal yield stress is affected by the total shear modulus of the alloy, atomic size mismatch and elastic modulus mismatch of the elements in the solid solution. These factors cause the yield stress plateau to occur at intermediate temperatures (400 ºC to 800 ºC) in HEAs. At these temperatures, the available thermal energy is insufficient to overcome the barriers to dislocation motion, resulting in limited plastic deformation. The thermally activated component of yield stress in RHEAs can be described using a phenomenological energy barrier governing screw dislocation motion, coupled with an Arrhenius-type relationship to capture the strain rate dependence. This model has successfully predicted the yield stress at various temperatures and observed the plateau as well. Overall, this model attributes the yield strength plateau solely to intrinsic factors like atomic mismatch, shear modulus and elastic modulus mismatch completely neglecting the extrinsic contributions like the effect of dislocations, multiple phases, solute clusters, and grain boundaries etc.

Even though there exist some analytical and atomistic models for the prediction of yield stress in RHEAs exist, there is a lack of comprehensive study in explaining the reason behind the yield strength plateau at medium and high temperature range based on the dislocation behavior and interactions with the solute atoms. In this study, two equimolar MoNbTaVW and MoNbTaW R-HEAs are developed and computationally studied using molecular dynamics simulations for mechanical strength at high temperature. The effect of any local-short range order around dislocation and thereby subsequently affecting the mechanical properties has not been studied with the dislocation theories. In this study, short range ordering in annealed RHEAs were studied using hybrid MC/MD simulations. Edge dislocations have been introduced in both random solid solution and MC evolved structures. Local elemental segregation and its impact on yield stress are examined in the following sections. The yield stress plateau is observed in our simulations and the reasons are elaborated in the Results and Discussion section.

## 2. Computational Methodology

An equimolar system of MoNbTaVW was created with supercell of 20 × 15 × 20 with the orthogonal axes extending along X = [1 1 1], Y = [-1 1 0] and Z = [-1 -1 2] directions using the package LAMMPS [36]. The lattice parameter was determined using classical molecular dynamics simulations with the LAMMPS package. At 0 K, energy minimization was performed starting from an initial guess of the lattice constant using the conjugate gradient algorithm. A single unit cell or a small supercell was created using the lattice and region commands, and the system was relaxed until the forces converged below the specified tolerance. The final box dimensions after minimization directly yielded the relaxed lattice parameter at 0 K. For finite temperature calculations, an isothermal-isobaric (NPT) ensemble was employed to allow the simulation cell to equilibrate at a target temperature (e.g., 300 K) and zero external pressure. The system was stabilized for a duration of 20 ps and over the equilibrated trajectory, the averaged box dimensions were taken to determine the lattice parameter. For systems constructed with multiple unit cells along each direction, the lattice constant was calculated by dividing the final box length by the number of unit cells. These methods provided temperature-dependent equilibrium lattice parameters for the system under study [37]. A lattice parameter of 3.203 Å at 0 K, 3.217 Å at 600 K, 3.22 Å at 1000 K, 3.231 Å at 1400 K, 3.235 Å at 1800 K and 3.244 Å at 2000 K was obtained. and the lattice parameter values almost match with the experimental value 3.19 Å [38]. The structures were made at different temperatures with these lattice parameters. In our work, we used Embedded Atom Type (EAM) potential for interactions between atoms in molecular dynamics simulations. By EAM potential, the total energy ($E_t$) of the system is given as [39]

$$E_t = \sum F_i \sum_{j \neq i} f_j(r_{ij}) + \frac{1}{2} \sum_{i,j(i \neq j)} \phi_{ij}(r_{ij}) \qquad (1)$$

where $F_i$ is the embedding function of the atom type at position $i$, $\phi_{ij}$ is the pairwise interaction function, f is the electron density function, $i$ and $j$ are the neighboring atoms within the mutual interaction range. The pairwise interaction of dissimilar atom types were calculated by the following relation.

$$\phi^{ab}(r) = \frac{1}{2}\left(\frac{f^b(r)}{f^a(r)}\phi^{aa}(r) + \frac{f^a(r)}{f^b(r)}\phi^{bb}(r)\right) \qquad (2)$$

where $\phi^{ab}(r)$ is the dissimilar interaction between atom types $a$ and $b$ at a distance of $r$, $f^a(r)$, $f^b(r)$ are their electron densities, $\phi^{aa}(r)$, $\phi^{bb}(r)$ are pair interaction energies of the pure element states of atom types $a$ and $b$ respectively. Electron density, embedding function and pairwise interaction data tables with respect to inter-atomic distance are estimated according to the methods used by Maiti et al. [40]. The component electron density of EAM potential is further expressed as

$$f(r) = f_e\left(\frac{r_e}{r}\right)^\beta ; f_e = \left(\frac{E_c}{S\Omega}\right) \qquad (3)$$

where $f_e$ is the scaling factor, $r_e$ is the first equilibrium neighbor, $E_c$ is the cohesive energy of element in consideration, $\Omega$ is the atomic volume, $\beta = 6$ and $S = 13.196$ for BCC lattice. The embedding function $F(\rho)$ of an element is expressed as

$$F(\rho) = -(E_c - E_f)\left[1 - \ln\left(\frac{\rho}{\rho^e}\right)^n\right]\left(\frac{\rho}{\rho^e}\right)^n ; n = \frac{1}{\beta}\left(\frac{9\Omega B - 15\Omega G}{E_c - E_f}\right)^{\frac{1}{2}} \qquad (4)$$

where $E_f$ is the unrelaxed vacancy formation energy, $\rho_e$ is the electron density from surrounding atoms in a relaxed lattice, B is the bulk modulus, G is the shear modulus, and $\Omega$ is the atomic volume. The pairwise interaction function $\Phi(r)$ is given as the spline function.

$$\Phi(r) = k_3\left[\frac{r}{r_e} - 1\right]^3 + k_2\left[\frac{r}{r_e} - 1\right]^2 + k_1\left[\frac{r}{r_e} - 1\right]^1 + k_0 \qquad (5)$$

where the constants $k_0$, $k_1$, $k_2$, and $k_3$ depend on the elastic constants of the element in consideration. Expressions of these constants $k_0$ to $k_3$ can be found in [40]. All the physical parameters required to obtain the different components of the EAM potential such as cohesive energy, lattice parameter, vacancy energy and second order elastic constants are taken from Maiti et al. [40].

The random alloy structures were replicated by 2 × 2 × 2 in each direction periodically to get a big supercell of 3,84,000 atoms. Then, an edge dislocation of ½ [111] was introduced in the middle of the structure along Z [-1 -1 2] and glide direction X [1 1 1] using the Osetsky method [41]. The big supercell has dimensions 225 Å, 134 Å, 210 Å along X, Y and Z directions.

Once the structures with edge dislocations were obtained, NPT simulations were carried out at 600 K, 1000 K, 1400 K, 1800 K, and 2000 K. Then the alloy structure created was subjected to hybrid MC/MD swaps at various temperatures. In hybrid MC/MD swaps, positions of two atoms were randomly chosen and swapped. Subsequently after the swap, the atomic positions were relaxed using the conjugate gradient method and the swap was accepted or rejected according to the Metropolis criterion [42] as mentioned below

$$Swap = \begin{cases} Accepted \,; \Delta U < 0 \\ Accepted \text{ with probability} \,; 0 < p < e^{-\frac{\Delta U}{k_B T}} \\ Rejected \,; p > e^{-\frac{\Delta U}{k_B T}} \end{cases}$$

As MC/MD swaps are computationally expensive for a large system with 381600 atoms, we divided the system into thin slices along dislocation of dimensions 225 Å × 134 Å × 20 Å comprising of around 8000 atoms. To get compositional variation along dislocation, hybrid MC/MD simulations were carried out in only one slice with z varying from 0 to 20 Å. After the swaps, the modified atom types in thin section were mapped back onto the corresponding atoms in the unrelaxed structure by matching similar atom IDs w.r.t their positions from dislocations. To extend the updated unrelaxed structure along the z-direction, the atomic data were first sorted with and z in ascending order. The atom types obtained from the swapped region were then systematically reassigned to the other thin slices, where the X and Y coordinates are similar, but the Z coordinates differ by an integer multiple of the slice thickness. By repeating this mapping procedure, the compositional variation observed in the thin section was periodically extended along the entire z-axis. The resulting structure was subsequently relaxed to eliminate residual stress, thereby preserving the dislocation periodicity and avoiding artificial strains. These systems were then energy-minimized with periodic boundary conditions applied along X and Z directions and a free boundary condition along Y direction, which resulted in a periodic array of dislocation (PAD) configuration [41]. This process was then repeated to 5 different configurations for averaging predictions of yield stress. The same boundary conditions were applied for hybrid MC/MD swaps as well. Similar process was repeated for MoNbTaW RHEA.

A pure shear stress was subsequently applied to the ($\bar{1}10$) glide plane in both the random and segregated structures, targeting the top 15 Å atoms, while the bottom 15 Å atoms were held fixed. The dislocation simulations were run at temperatures at 600K, 1000K, 1400K, 1800K and 2000K using NVE ensemble in LAMMPS. The dislocation motion and its interactions with atoms surrounding the dislocation core were observed. The shear stress level was taken as stress at which the dislocation started to move for around 110 Å in 65 ps as taken in earlier work from our group [43]. Dislocations in the systems were identified using the Common Neighbor Analysis (CNA) and dislocation analysis tools available in OVITO [44]. The Dislocation analysis (DXA) algorithm was employed to identify the dislocation lines, and its movements were observed under the applied shear stress [45].

## 3. Results and Discussion

### 3.1 Microstructure and Dislocation

Initially in an equimolar RHEA MoNbTaVW the atoms were randomly assigned to five different types. An edge dislocation was introduced in the middle of the simulation cell using Osetsky method [41] as shown in Fig. 2. The MC/MD swaps were attempted for 5 Monte Carlo swaps per atom at various temperatures from 600K to 2000K for all the 5 configurations for the thin section of atoms of dimensions 225 Å × 134 Å × 20 Å. The energy of the system was observed to converge after around 2 atomic swaps per atom for temperatures 1000 K, 1400 K, 1800 K and 2000 K whereas for 600 K, the energy started converging after 5 atomic swaps per atom as shown in Fig.3. At all temperatures, the energy converges within these five swaps per atom. The OVITO images of Mo, Nb, Ta, V, and W before and after 5 swaps per atom are shown in Fig.4.

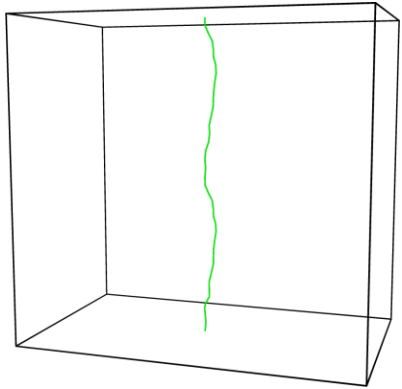

Fig.2 Edge dislocation in MoNbTaVW RHEA simulated using Osetsky method.

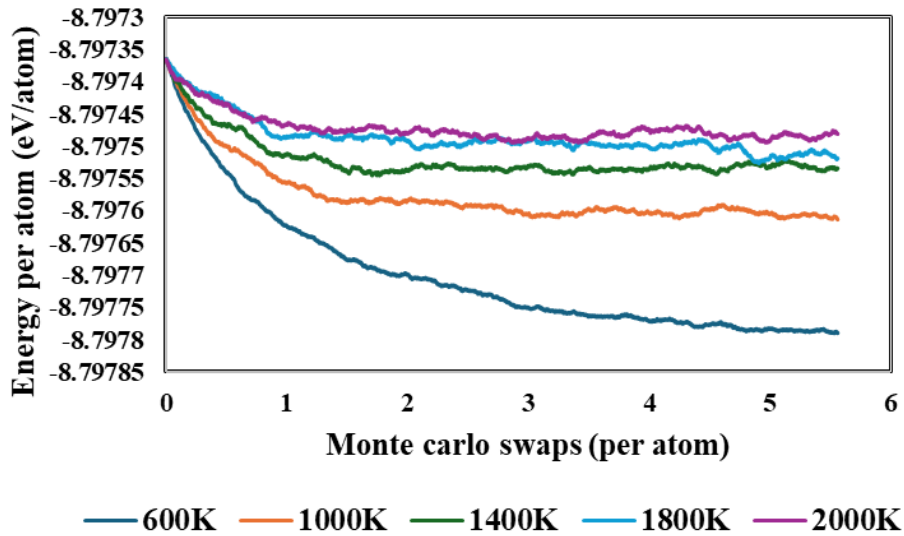

Fig.3 Variation in system energy (per atom) with Monte Carlo swaps (per atom) for MoNbTaVW RHEA.

From Fig.4 it can be found that some elements, especially Nb and V are getting segregated on the compressive and tensile areas of the edge dislocation consistent with their atomic sizes [46] because Nb has an atomic radius of 1.43 Å and V has an atomic radius of 1.34 Å. Typically elements with smaller atomic radii get segregated on the compressive part of the edge dislocation and vice versa [46]. This results in short-range ordering (SRO) around dislocation core region. The SRO around the dislocation core has been observed experimentally by Atom Probe Tomography (APT) in other RHEAs such as AlTiNbZrTa [47] as well as in high alloy and low alloy steels [48][49]. In our edge dislocation configuration, the extra half-plane lies below the slip plane, resulting in a stress field where the tensile region is above and the compressive region is below. In contrast, Mo and W are observed to move away from the dislocation core after MC/MD swaps. This repulsion may be attributed to their moderate atomic size mismatch with the matrix, resulting in weak or repulsive elastic interactions with the dislocation stress field. For MoNbTaW alloy (without V), Mo and W were moving away from the dislocation core, Nb was segregated on the compressive area of edge dislocation and Ta did not show any significant variation which is represented in Fig.5.

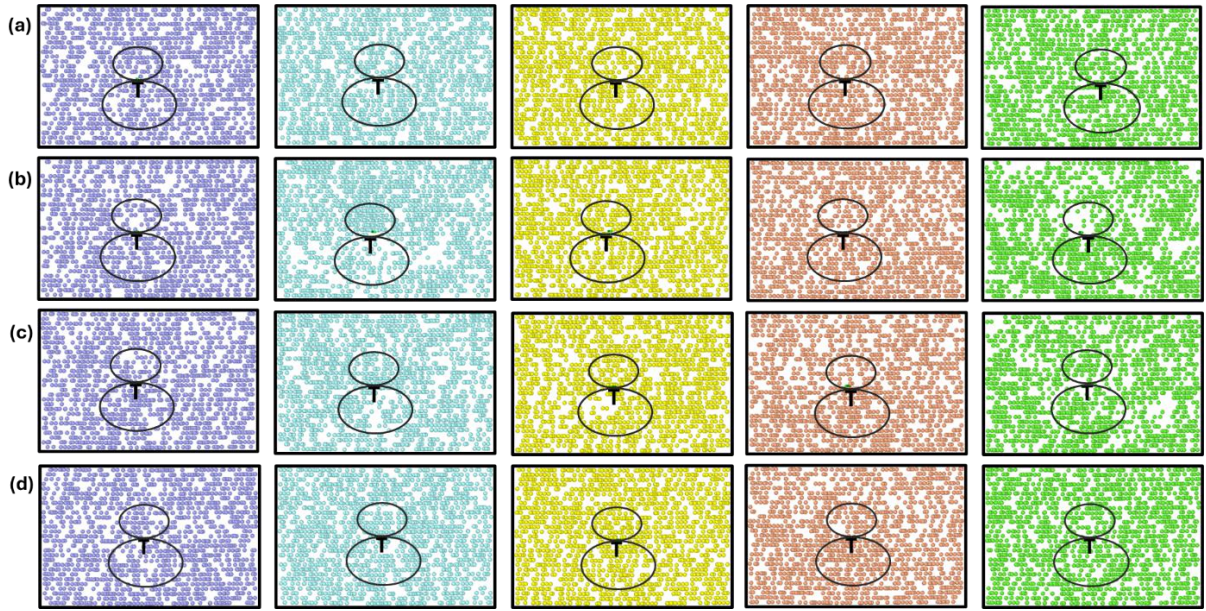

Fig. 4 Cross-section of MC-evolved structure around the dislocation core with different colored atom types in MoNbTaVW: Mo (purple), Nb (sky blue), Ta (yellow), V(orange), and W (green). (a) Initial structure at 600 K. Segregated structures after 5 swaps per atom at (b) 600 K, (c) 1000 K, and (d) 2000 K. Black circles in the snapshot represent the segregation behavior of atoms in alloys.

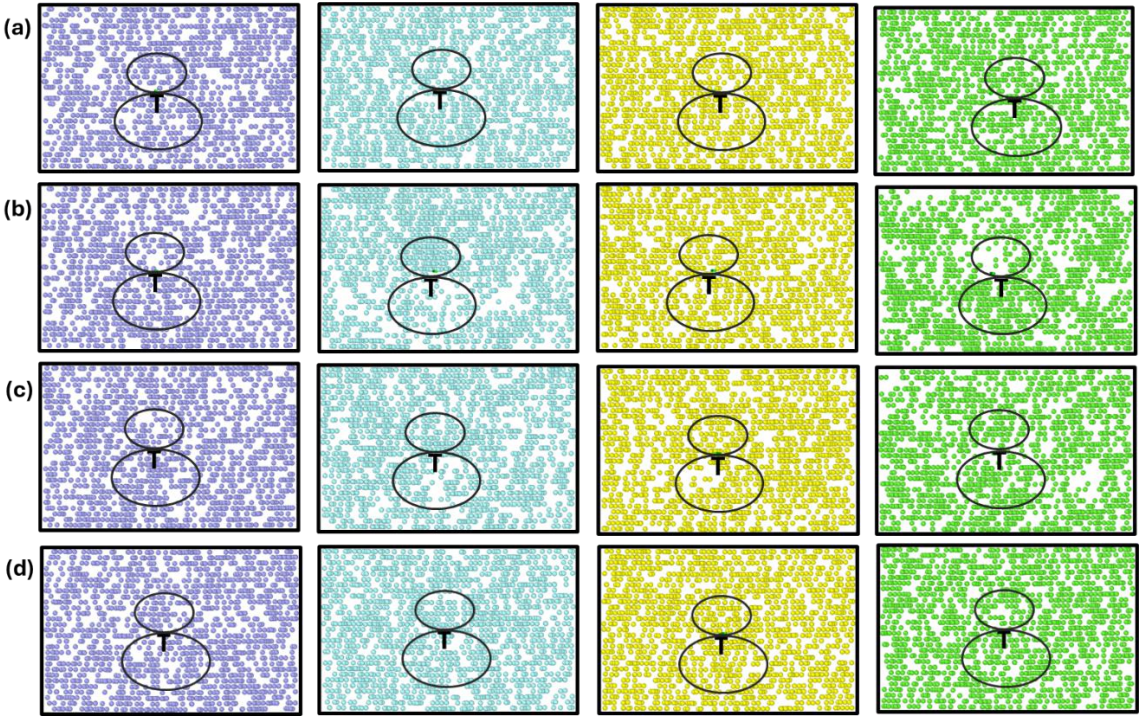

Fig.5. Cross-section of MC-evolved structure around the dislocation core with different colored atom types in MoNbTaW: Mo (purple), Nb (sky blue), Ta (yellow), and W (green). (a) Initial structure at 600 K. Segregated structures after 5 swaps per atom at (b) 600 K, (c) 1000 K, and (d) 2000 K. Black circles in the snapshot represent the segregation behavior of atoms in alloys.

### 3.2 Diffusivity and residence time calculations:

Refractory high entropy alloys (RHEAs) have been observed to exhibit sluggish diffusion in bulk [50-52]. Bulk diffusivity in solids is governed by the ease with which the atoms traverse through the bulk. The vacancy formation energy in the bulk lattice is generally high due to the dense atomic packing. Interestingly, it is observed that vacancies are thermodynamically favored in the compressive region of dislocation stress field around the dislocation core [53-55]. Vacancies on dislocation lines can be considered as fundamental super-jogs which help in pinning the edge dislocations, thereby contributing to the improved strength even at intermediate and high temperatures. The diffusivity of atoms in solids (bulk or dislocation core) can be calculated as

$$D = D_0 \exp\left(-\frac{\Delta H}{k_B T}\right) \qquad (6)$$

where $D_0$ is the pre-exponential factor, $\Delta H$ is the activation enthalpy, $k_B$ is the Boltzmann constant ($8.617 \times 10^{-5}$ eV/K), T is the temperature in K.

$\Delta H$ can be calculated as summation of vacancy formation energy and migration barrier.

$$\Delta H = E_f + E_m \qquad (7)$$

$E_f$ determines how many vacancies can be formed and $E_m$ determines how easily atoms can jump to nearby vacancies.

$D_0$ can be given as [50]

$$D_0 = \frac{1}{6} z f v_0 b^2 \qquad (8)$$

where $z = 8$ which is the coordination number for 1st nearest neighbor in bcc lattices, $f = 0.75$ is a correlation factor, $b$ is the jump distance of 1st nearest neighbor which can be given as $b = \frac{\sqrt{3}}{2} a_0$ and $a_0$ is the lattice parameter. The frequency $v_0$ is Debye frequency which can be given as

$$v_0^3 = 6\pi^2 n c^3 \qquad (9)$$

where n is the atomic density, $c$ is the sound's velocity in the material, expressed as following:

$$n = \frac{2}{a_0^3} \qquad (10)$$

$$c = \sqrt{\frac{K}{\rho}} \qquad (11)$$

where $K$ is the shear modulus, $\rho$ is the mass density computed using the following equation.

$$\rho = \frac{2m}{N_A a_0^3} \qquad (12)$$

where $m$ is the atomic weight, $N_A$ is Avogadro's constant, and 2 represents the number of atoms in a unit cell of bcc structure.

The pre-exponential factor $D_0$ is estimated as $2.96 \times 10^{-6}$ $m^2/s$ for MoNbTaW RHEA and $3.198 \times 10^{-6}$ $m^2$/s for MoNbTaVW using equations (8-12) [50].

The formation and migration energy of vacancy in the bulk lattice and dislocation core of MoNbTaW were calculated using Density Functional Theory simulation in the literature, as mentioned in Table 1 [53]. From these results, it can be inferred that vacancy formation energy value around the dislocation core is much less than the vacancy formation energy in the bulk indicating that it is easier to form a vacancy around the dislocation core than in the bulk. This can be attributed to the fact that dislocations are the region of local atomic disorder due to which the energy cost of removing atoms is less. Additionally, the migration energy barrier around the dislocation core is much less than the bulk, indicating that atoms can migrate into the vacancies more easily around the dislocation core as compared to the bulk lattice.

The vacancy formation energy ($E_f$) and migration energy ($E_m$) for MoNbTaVW RHEA in bulk lattice has been estimated using first-principal calculations to be 3.3 eV and 1.258 eV, respectively, by Byggmaster et. al [46]. However, these values are expected to significantly reduce in the edge dislocation due to ease of diffusion of atoms around the dislocation due to the local atomic distortions. To estimate the core-level vacancy energetics in MoNbTaVW, we utilized the known core-to-bulk energy ratios from a chemically similar RHEA system, MoNbTaW, for which both core and bulk values are available which is mentioned in Table 1. From these values, we determine the ratios of $E_f$(core)/$E_f$(bulk) and $E_m$(core)/$E_m$(bulk) as 0.42 and 0.187.

A similar ratio of energetics in formation and migration for core to bulk $E_f$(core)/$E_f$(bulk) for other BCC metals such as Mo (0.36), Nb (0.36), Fe (0.44) has been observed by researchers [58-60]. Assuming a similar reduction in energy at the dislocation core for MoNbTaVW due to analogous chemical and structural environments, we applied the same ratios to estimate the core values which are given in Table 1. The obtained estimates provide an adequate representation of the vacancy energetics at the dislocation core in MoNbTaVW.

Table 1. $E_f$ and $E_m$ values of core and bulk for MoNbTaVW and MoNbTaW

| Energies (eV) | MoNbTaW | MoNbTaVW |
|---|---|---|
| $E_f$ (bulk) | 2.48 | 3.3 |
| $E_m$ (bulk) | 1.71 | 1.258 |
| $E_f$ (edge dislocation core) | 1.05 | 1.386 |
| $E_m$ (edge dislocation core) | 0.32 | 0.235 |

Using the predetermined $D_0$ and $\Delta H$ values, the diffusivities across the dislocation core of MoNbTaW and MoNbTaVW alloys were calculated at various temperatures, as presented in Fig 6. Also, it was observed that diffusivity of atoms is $10^{14}$ times higher in dislocation core region as compared to bulk lattice. This proves that atoms can diffuse at a faster rate in dislocation region and can form an SRO in these alloys as suggested using hybrid MC/MD simulations by Wang et.al and Zhao et.al [56-57]. Also, it is observed that at the intermediate temperatures (1200 K – 1400 K) the dislocation core diffusivity value is changing slowly which indicates that at those temperatures, there is no significant change in diffusion of atoms.

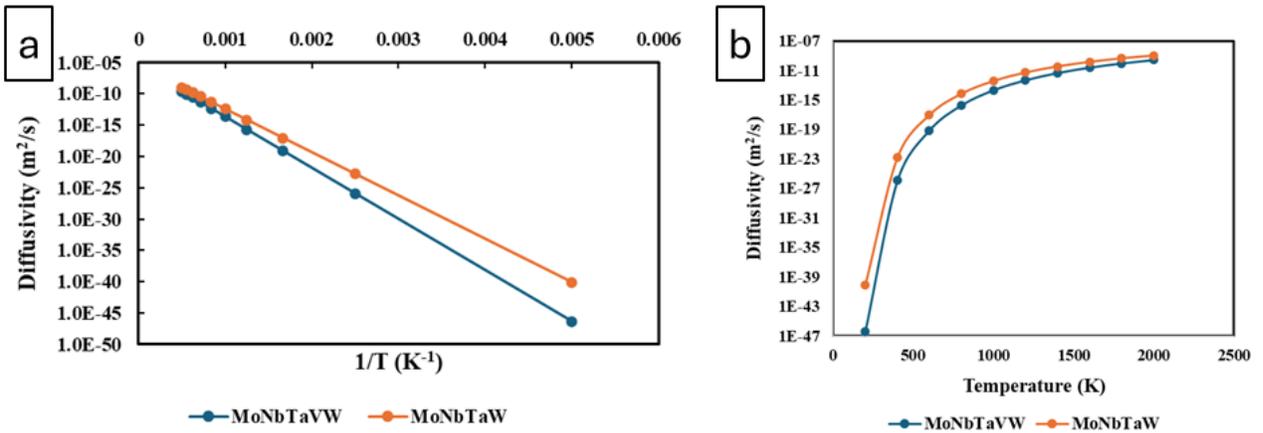

Fig.6 Diffusivity values for MoNbTaVW and MoNbTaW (a) w.r.t 1/T (K$^{-1}$) (b) w.r.t T (K)

### 3.3 Residence time and critical swaps:

The residence time or waiting time ($t_w$) is defined as the period during which mobile dislocations are temporarily halted by solute clusters. To overcome such halts, thermal activation or stress is required which is associated with energy barriers [61]. Thus, the waiting time is attributed to being influenced by temperature, strain rates and applied stress. The waiting time can be estimated in various ways with respect to the dislocation density, burgers vector and strain rate [61]. A common method for estimating residence time involves dividing the average velocity of the dislocation by the solute pinning distance. Solute pinning distance refers to the average spacing between solute atoms that interact with dislocation segment and hinder its motion. The dislocation velocity can be determined using the Orowan equation which is given as [61].

$$\dot{\varepsilon} = \frac{\rho b v}{2} \quad (13)$$

where $\dot{\varepsilon}$ is the strain rate, $\rho$ is the dislocation density and $v$ is the dislocation velocity. ½ is schimd factor between shear strain and tensile strain in polycrystalline materials [61].

Edge dislocations in typical MoNbTaVW and MoNbTaW RHEAs developed experimentally, have a dislocation density in the range of $10^{12} \ m^{-2}$ [62]. For dislocation motion in experiments, a typical strain rate of $10^{-3} \ s^{-1}$ is applied and critical stress is calculated. Burgers vectors of edge dislocations were estimated by lattice constant derived from relaxed structures of molecular dynamics simulations conducted earlier in this work. This was determined to be 2.794 Å and 2.763 Å for MoNbTaW and MoNbTaVW alloys, respectively. The pinning distance was assumed to be approximately 2.7 Å as the average distance between two atoms in dislocation core region in RHEAs. This was calculated by analyzing a region within 10 Å around the dislocation core in the segregated structure using the Coordination Analysis modifier in OVITO [63]. The resulting radial distribution function (RDF) plots showed a prominent first peak at 2.7 Å as the pair separation distance for all the solute pairs, which was interpreted as the characteristic closest possible solute pinning distance. The dislocation waiting time is given as

$$t_w = \frac{d}{v} \quad (14)$$

where $t_w$ is the waiting time (s), $v$ is the dislocation velocity (m/s), $d$ is the pinning distance (m).

Using these experimentally derived values for RHEAs in our study, dislocation velocity was calculated as represented in Table 2. Accordingly, the dislocation waiting time for both the RHEAs is calculated and given in Table 2.

Table 2. Dislocation velocity and waiting time of MoNbTaVW and MoNbTaW

| Alloy | Dislocation velocity (m/s) | Waiting time (s) |
|---|---|---|
| MoNbTaW | 0.715E-05 | 3.78E-05 |
| MoNbTaVW | 0.723E-05 | 3.73E-05 |

The atoms in the dislocation core region can diffuse only within this waiting time. The variation in concentration can be obtained by the critical number of total swaps allowed/possible within this waiting time. The critical number of total swaps ($N$) is equivalent to successful atomic jumps, which can be estimated as following [64]

$$N = \frac{\overline{R_n^2}}{\lambda^2} \qquad (15)$$

where $\overline{R_n^2}$ is the mean square displacement of atoms, and $\lambda$ is the jump distance which is around 2.763 Å for MoNbTaVW and 2.794 Å for MoNbTaW.

The mean square displacement of atoms can be estimated as [64]

$$\overline{R_n^2} = \sqrt{2Dt_w} \qquad (16)$$

where $D$ is the diffusivity along the dislocation core, $t_w$ is the waiting time.

We took a region of approximately 8000 atoms around the dislocation core to be swapped in hybrid MC/MD simulations. For such a region, the mean square displacement of atoms, and critical number of swaps allowed per atom in hybrid MC/MD simulation at each temperature were calculated, as mentioned in Table 3. From Table 3, it is interpreted that at intermediate temperatures analogous to the diffusivities the critical number of swaps is rising slowly, indicating that at the intermediate temperatures and rapidly above 1400K.

Table.3 Number of critical swaps per atom and length of segregation vs temperatures

| Temperature (K) | Critical number of swaps per atom | | Mean square displacement of atoms ($\overline{R_n^2}$) (Å²) | |
|---|---|---|---|---|
| | MoNbTaVW | MoNbTaW | MoNbTaVW | MoNbTaW |
| 0 | 0.00 | 0.00 | 0.00E+00 | 0.00E+00 |
| 200 | 0.00 | 0.00 | 5.81E-16 | 8.18E-13 |
| 400 | 0.00 | 0.00 | 9.47E-06 | 3.50E-04 |
| 600 | 0.00 | 0.00 | 2.40E-02 | 2.64E-01 |
| 800 | 0.00 | 0.00 | 1.21E+00 | 7.23E+00 |
| 1000 | 0.00 | 0.04 | 1.27E+01 | 5.28E+01 |
| 1200 | 0.06 | 0.63 | 6.09E+01 | 1.99E+02 |
| 1400 | 0.57 | 4.15 | 1.87E+02 | 5.12E+02 |
| 1600 | 3.03 | 17.17 | 4.32E+02 | 1.04E+03 |
| 1800 | 11.20 | 51.80 | 8.31E+02 | 1.81E+03 |
| 2000 | 31.84 | 125.30 | 1.40E+03 | 2.81E+03 |

### 3.4 Edge dislocation and yield strength plateau in MoNbTaVW and MoNbTaW RHEAs

Edge dislocations were introduced using the Osetsky method [41]. Dislocations were relaxed to achieve equilibrium atomic positions, allowing the dislocation core to separate out distinctly following a Conjugate Gradient (CG) relaxation. This dislocation insertion is done in all the systems from random solid solution (0 MC swaps) to 5 MC swaps for both the alloys. Fig. 7 illustrates selected evolved structures (at 0 and 5 Monte Carlo swaps) containing edge dislocations aligned along the ($\bar{1}10$) glide planes, captured at various stages of their glide motion under applied stress. Figures 7a1–a3 correspond to the initial configuration at 0 Monte Carlo (MC) swaps, while Figs. 7b1–b3 represent the structure after 5 MC swaps at 600K. In these visualizations, blue atoms indicate regions conforming to the local BCC crystal lattice, whereas white atoms signify deviations from this lattice, as identified using the Common Neighbor Analysis (CNA) feature in OVITO [44]. The dislocation line is discernible as a continuous arrangement of white atoms forming a curved path, with the dislocation core highlighted by a green line generated through Dislocation Extraction Algorithm (DXA) analysis. Upon application of shear stress, the dislocation line begins to glide from the right to the left side of the simulation box. Due to the Periodic Array of Dislocation (PAD) configuration, the dislocation exits the box on the right and re-enters from the right, consistent with the imposed periodic boundary conditions. As shown in Figures 7b2, b3, the dislocation initially adopts a zigzag configuration upon introduction and relaxation, with its core exhibiting multiple kinks. This localized waviness and kinked structure of the dislocation core has also been observed in other dislocation-based simulations involving High Entropy Alloys (HEAs) and multi-element Compositionally Complex Alloys (CCAs) [17][43][65]. Such rough and undulating core morphology is attributed to nanoscale variations in the local energy landscape, which enables the dislocation to navigate and settle along a minimum energy path within that landscape [17][22][65].

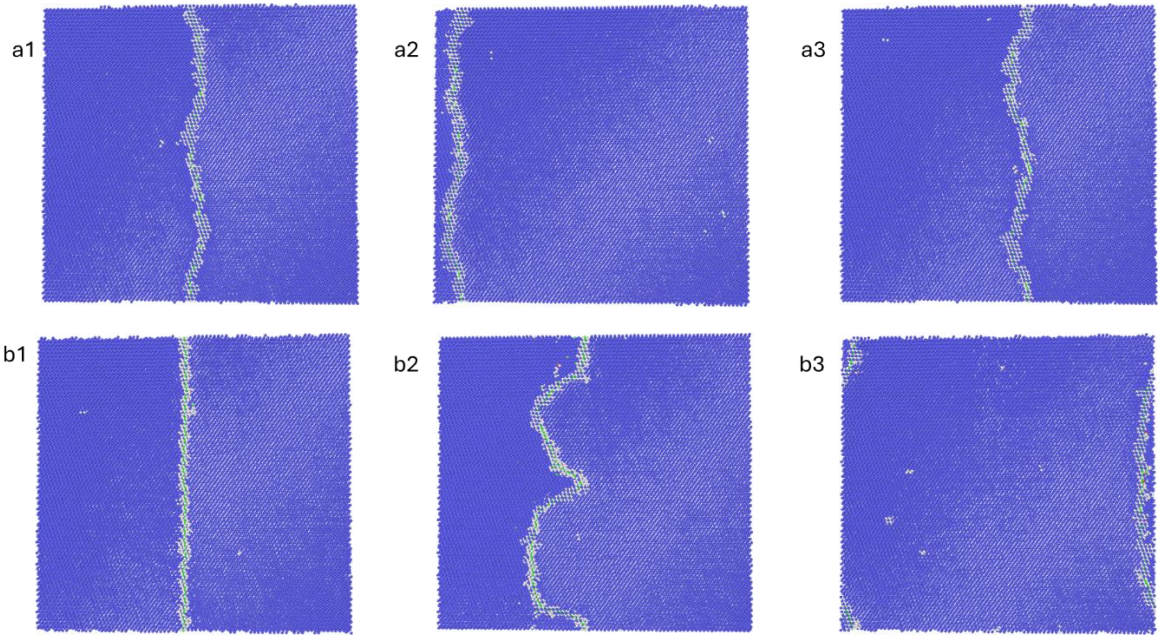

Fig.7: Movement of edge dislocation (represented by green color) from right to left under the application of shear stress. (a1, a2, a3) are snapshots of dislocation position in system for 0 MC swap at 600K after 0 ps, 40 ps, 60 ps respectively. (b1, b2, b3) are snapshots of dislocation position in system for 0 MC swap at 600K after 0 ps, 50 ps, 80 ps respectively.

Initially, shear stress was applied at 600 K. Beyond a certain threshold of the critically resolved shear stress (CRSS), the dislocation began to glide toward the right. During its motion, several segments of the dislocation appeared to become locally pinned, likely due to the presence of multiple energy minima—an effect also reported in previous studies on HEAs [22][65]. The shear stress was incrementally increased in steps, and the CRSS was identified as the stress level at which the dislocation could overcome these pinning sites and traverse the entire simulation cell along the [111] direction. At this estimated CRSS, the dislocation advanced approximately 110 Å over a simulation time of 35 ps approximately with the velocity of 150 m/s. A similar approach for CRSS estimation using constant stress application has been adopted in earlier work. [22][43][65] From CRSS, yield strength can be estimated from Taylor's relation as

$$\sigma_y = M * \tau \tag{17}$$

where $M$ is the Taylor factor, which is 3.067 for polycrystalline materials, $\tau$ is the Critically resolved shear stress (CRSS) [17].

As seen in Figures 7a1–a3, the edge dislocation in the random RHEA leaves behind atoms that deviate from the local BCC lattice as it moves from the center toward the left. These residual atoms, referred to as 'debris,' have also been observed in other HEA systems [65]. The persistent waviness of the dislocation line, even within the random alloy structure, and the observed local pinning effects highlight the highly rugged atomic energy landscape. This ruggedness arises from statistical variations in the local chemical environment, which the dislocation must navigate often by leaving behind vacancies and interstitial-type debris [22, 65-66]. Similar debris formation along the dislocation glide path is observed in the Monte Carlo (MC) evolved structures at 600 K, as shown in Figures 7b1–b3 for both 0 and 5 MC swaps. The amplitude of the local waviness of the dislocation line appears to increase from 0 to 5 MC swaps, likely due to enhanced pinning effects arising from chemical short-range order around the initial dislocation line [43].

The solid solution strengthening of random alloys can be represented as the following relation of shear strength ($\tau$ )[62]

$$\tau = \tau^* + \tau_G + \tau_{size} \tag{18}$$

where $\tau^*$ is the friction resistance of lattice for dislocation movement, $\tau_G$ is the Taylor hardening component and $\tau_{size}$ is the grain-size dependent strengthening. In this study, as well as in the corresponding yield experiments conducted on the MoNbTaVW and MoNbTaW alloys, the random solid-solution system was not subjected to prior plastic deformation or work hardening. Consequently, the material lacks any pre-existing dislocation density, and the Taylor hardening component ($\tau_G$), which arises from dislocation interactions, is effectively zero. Experimental studies on the compositionally related MoNbTaVW and MoNbTaW alloys have established a grain size-dependent strengthening relationship represented as $\frac{240}{\sqrt{D}}$ MPa, where $D$ is the grain size in micrometers. For MoNbTaVW and MoNbTaW the grain size is about 80 μm and 200 μm, respectively [38]. The calculated $\tau_{size}$ values of approximately 26 MPa and 16 MPa are relatively minor and can be considered negligible. From the above discussions, it is interpreted that shear strength is predominantly influenced by dislocation lattice resistance $\tau^*$.

Conventional BCC alloys with a single principal element primarily attribute their strengthening mechanisms to screw dislocation activity and the kink-pair nucleation process [17][39][65]. However, in multi-principal element alloys, dislocation motion is significantly impeded by a highly heterogeneous atomic environment and locally fluctuating energy landscapes, which introduce substantial activation barriers [17][22][65].

To visualize the plateauing effect in RHEAs, we applied shear stress on the edge dislocations at the MC evolved structures at the critical swaps calculated in earlier section at various temperatures.

Given that the critical number of swaps significantly exceeds the threshold for both alloys beyond 1800K, we have taken 5 as the critical number of swaps for the range of 1800K to 2000K for both alloys. After 5 swaps, the system's energy was observed to converge with minimal variation in segregation.

Subsequently, we determined the Critically Resolved Shear Stress (CRSS) at each temperature for both MoNbTaVW and MoNbTaW, corresponding to the identified critical number of atomic swaps for cross core motion which is shown in Fig.8. It is evident from Fig.8 that the critical shear stress required for the dislocation motion due to dislocation core diffusion increases to 1000K and reaches a plateau at 1400K and then starts declining at 1800K. This saturation at the intermediate temperatures implies that all the available diffusion-driven core solute strengthening mechanisms have been achieved, no further solute drag enhancement is possible beyond this point. This is the reason for the exceptional yield strength plateauing at intermediate temperatures. This phenomenon is called cross core diffusion via Dynamic strain ageing (DSA) mechanisms, observed in RHEAs as serrated stress strain curves in other literatures as well [25]. However, the extent of saturation observed in CRSS may be little higher due to a higher dislocation density ($\sim 3 \times 10^{15}\ m^{-2}$) in our systems than typically observed in experimental studies. The choice of higher dislocation density was made to balance the computational requirements of our simulations for limited system sizes.

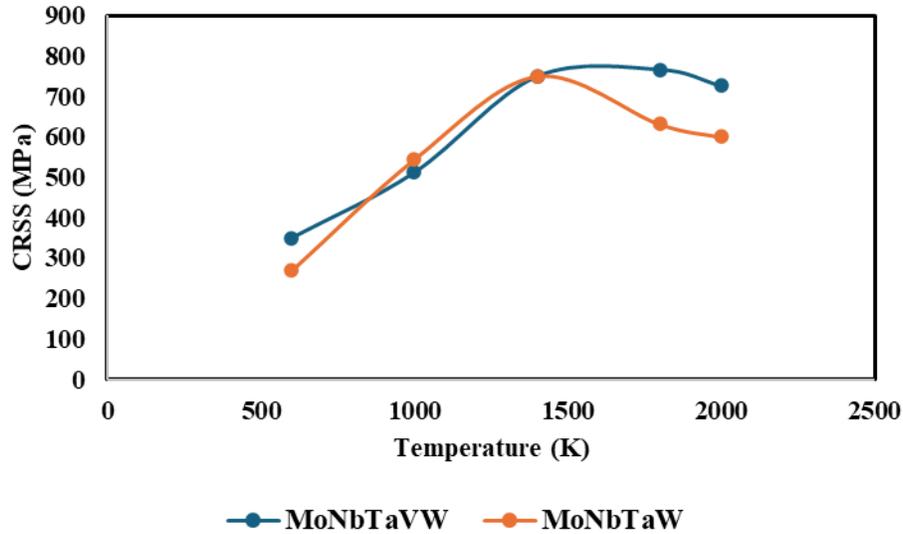

Fig.8 CRSS (MPa) vs Temperature (K) for MoNbTaVW at the critical swaps.

Similarly, we applied shear stress even on edge dislocations of random solid solution alloys - 0 MC swaps at different temperatures and found similar trends to literature as shown in Fig.9. Interestingly, we find that the yield stress remains relatively constant between 1100 K and 1500 K, after which it begins to decline little with increasing temperature. This behavior aligns with observations from literature on MoNbTaVW and MoNbTaW alloys for high temperature strengths [15][17]. In this case the yield strength plateau can be attributed to athermal stress mechanisms as mentioned by [15][28][29]. Although there is little explicit experimental validation of the existence of athermal yield stress, it is not good to disregard it completely. As we had already discussed in the earlier section [29], the athermal component of the stress field cannot be overcome by dislocations by thermal fluctuations. This could be due to the longer-range stress field with larger energy barriers coming from the atomic size mismatch, elastic modulus mismatch, arising due to the inherent compositional fluctuations in the material's structure [28]. Ultimately, dislocation motion is prevented especially at intermediate temperatures when the thermal energy is not enough to bypass both the athermal and cross core diffusion barrier. This distinctive thermal behavior sets

RHEAs apart from conventional superalloys, highlighting their potential for reliable performance in extreme high-temperature environments.

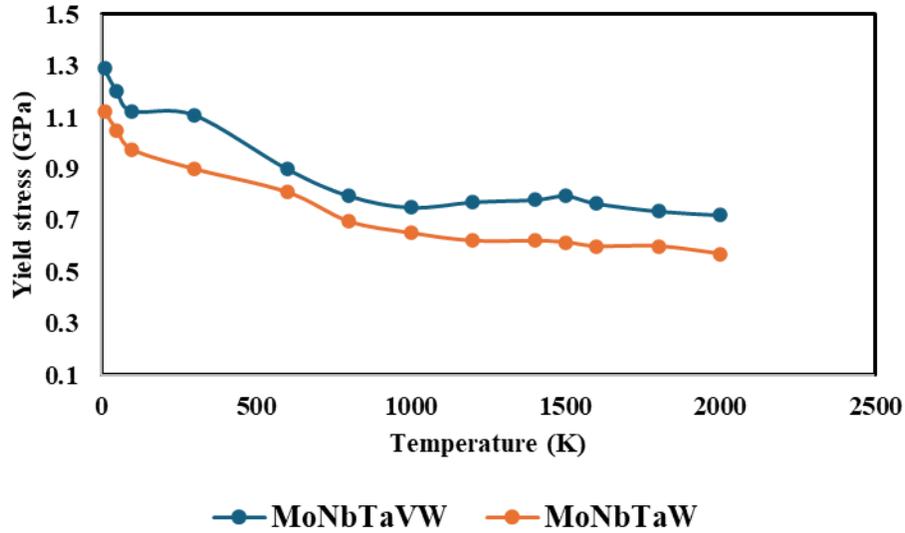

Fig. 9 Yield stress (GPa) vs Temperature (K) for MoNbTaVW and MoNbTaW random solid solution.

## 4. Conclusions

It can be interpreted from the whole simulation study of MoNbTaVW and MoNbTaW RHEAs that many of thermodynamic, microstructural, strengthening and dislocation related properties can be obtained from computational studies. Edge dislocations were created in both the random solid solutions and in MC evolved structures. The number of effective and critical MC swaps around dislocations were determined based on the diffusivities at various temperatures. The CRSS obtained by observing the movement of dislocation in the random solid solution alloy resembled qualitatively and quantitatively with the experimental values. The dislocation moved under applied shear stress with many local pinning and nanoscale detrapping mechanisms leaving behind debris of vacancy or interstitials. The yield strength plateau observed in both the alloy systems MoNbTaVW and MoNbTaW within the temperature range 1100K-1500K indicates a deviation from thermal softening usually seen at high temperatures. This behavior may arise due to a combination of athermal stress contributions and dynamic strain ageing (DSA) effects. Athermal mechanisms attribute this anomaly to solid solution strengthening which is inherently present in RHEAs due to its complexity and local structure. This type of strengthening impedes the dislocation mobility irrespective of thermal effects due to variation in local stress fields and statistical compositional fluctuations. On the other hand, DSA mechanisms arise due to the "cross core diffusion" – local diffusion of solute atoms only across the dislocation core. This results in time and temperature-dependent strengthening, which saturates once all cross-core motion is complete, after which the typical decrease in strength with increasing temperature resumes. A similar trend was observed in our simulations where stress required to move the dislocation gets saturated after 1400K. In structures where atomic diffusion is allowed, stress required to move the dislocation gets saturated – consistent with DSA behavior. Interestingly, even in random solid solution configurations with no atomic swaps (i.e., no diffusion), the yield strength plateau is still evident, which could be due to the athermal stress barriers arising from solid solution local structural variation effects. The concurrent appearance of the plateau in both diffused-dislocation and random configurations highlights the interplay of thermally activated and athermal mechanisms. Therefore, yield strength retention in this intermediate temperature range is likely governed by both DSA and inherent athermal stress contributions.

**Authors contribution**

SM ideated, planned and guided the work. SAS did developmental and simulation work, AG did analysis on diffusion related phenomenon, BR contributed to overall guidance of this work.

**Data availability**

Data available on request

**Conflicts of interest**

The authors declare that they have no known competing financial interests or personal relationships that could have appeared to influence the work reported in this paper.